\documentclass[12pt]{iopart}
\usepackage{iopams}
\begin{document}
\letter{On the self-similarities of the Penrose tiling}
\author{Nicolae Cotfas}
\address{Faculty of Physics, University of Bucharest,
PO Box 76-54, Postal Office 76, Bucharest, Romania, 
E-mail address: ncotfas@yahoo.com}
\jl{1}
\begin{abstract}
We show that the well known two-dimensional Penrose tiling admits an infinite number of
independent scaling factors and an infinite number of inflation centers.
\end{abstract}
\maketitle

\vspace{3mm}

The relations 
\begin{equation}\label{rep1} 
a(\alpha ,\beta )=(c\alpha -s\beta ,s\alpha +c\beta )\qquad \qquad  b(\alpha ,\beta )=(\alpha ,-\beta ) 
\end{equation} 
where  
$c=\cos (\pi /5)=(1+\sqrt{5})/4,$  
$s=\sin (\pi /5)=\sqrt{10-2\sqrt{5}}/4$ 
define the usual two-dimensional representation of the dihedral group  
\[ D_{10}=\left< a,\; b\ \left|\ a^{10}=b^2=(ab)^2=e\right. \right>.\] 
Let $\mathbb{E}_5=(\mathbb{R}^5,\langle ,\rangle )$ be the usual 
five-dimensional Euclidean space, $\varepsilon _1=(1,0,...,0),\ 
\varepsilon _2=(0,1,0,...,0),\ ...,\ \varepsilon _5=(0,...,0,1)$ 
be the vectors of the canonical basis of $\mathbb{E}_5$, and let
$c'=\cos (2\pi /5)=(\sqrt{5}-1)/4,$ 
$s'=\sin (2\pi /5)=\sqrt{10+2\sqrt{5}}/4.$ 
The $D_{10}$-cluster generated by $(1,0)$ 
\[ {\mathcal C}=D_{10}(1,0)=\{ e_1,e_2,e_3,e_4,e_5,-e_1,-e_2,-e_3,-e_4,-e_5\} \] 
where $e_1=(1,0),$ $e_2=(c',s'),$ $e_3=(-c,s),$ $e_4=(-c,-s),$ $e_5=(c',-s')$ 
is formed by the vertices of a regular decagon.

The action of $a$ and $b$ on 
${\mathcal C}$ is described by the signed permutations
\[
  a=\left( \begin{array}{rrrrr}
                e_1 & e_2 & e_3 & e_4 & e_5\\
                -e_4 & -e_5 & -e_1 & -e_2 & -e_3 
                \end{array} \right)
\qquad b=\left( \begin{array}{rrrrr}
                e_1 & e_2 & e_3 & e_4 & e_5\\
                e_1 & e_5 & e_4 & e_3 & e_2  
                \end{array} \right) 
\]
and the corresponding transformations $a,\ b:\mathbb{E}_5\longrightarrow \mathbb{E}_5$
\[
 a=\left( \begin{array}{rrrrr}
                \varepsilon _1 & \varepsilon _2 & \varepsilon _3 & 
                \varepsilon _4 & \varepsilon _5 \\
                -\varepsilon _4 & -\varepsilon _5 & -\varepsilon _1 & 
                -\varepsilon _2 & -\varepsilon _3
                \end{array} \right)
\qquad  b=\left( \begin{array}{rrrrr}
     \varepsilon _1 & \varepsilon _2 & \varepsilon _3 & \varepsilon _4 & 
     \varepsilon _5\\
     \varepsilon _1 & \varepsilon _5 & \varepsilon _4 & \varepsilon _3 & 
     \varepsilon _2
                \end{array} \right) 
\]
generate the orthogonal representation of $D_{10}$ in $\mathbb{E}_5$ 
\begin{equation} \begin{array}{l} 
a(x_1,x_2,x_3,x_4,x_5)=(-x_3,-x_4,-x_5,-x_1,-x_2)\\ 
b(x_1,x_2,x_3,x_4,x_5)=(x_1,x_5,x_4,x_3,x_2). 
\end{array}  
\end{equation} 
 
The vectors 
$u_1=\varrho (1,c',-c,-c,c'),$ $u_2=\varrho (0,s',s,-s,-s'),$ 
where $\varrho =\sqrt{2/5},$ form an orthonormal basis of the  
$D_{10}$-invariant subspace \cite{C2}
\begin{equation} 
 E= 
\left\{ (<r,e_1>,<r,e_2>,...,<r,e_5>)\ 
|\ \ r\in \mathbb{E}_2\ \right\} 
\end{equation} 
and the isometry (which is an isomorphism of representations) 
\begin{equation} 
\mathcal{I}:\mathbb{E}_2\longrightarrow E: r\mapsto 
(\varrho <r,e_1>,\varrho <r,e_2>,...,\varrho <r,e_5>) 
\end{equation} 
with the property $\mathcal{I}(\alpha ,\beta )=\alpha u_1+\beta u_2$ allows us  
to identify the physical space $\mathbb{E}_2$ with the subspace $E$ of $\mathbb{E}_5$. 
The matrices of the orthogonal projectors 
$\pi ,\, \pi ^\perp :\mathbb{E}_5\longrightarrow \mathbb{E}_5$
corresponding to $E$ and 
\begin{equation}
E^\perp =\{ x\in \mathbb{E}_5\ |\ \langle x,y\rangle =0 \
{\rm for\ all\ } y\in E\}
\end{equation}
in the basis $\{ \varepsilon _1,\varepsilon _2,\varepsilon _3,
\varepsilon _4,\varepsilon _5\}$ are
\begin{equation} 
\pi =\mathcal{A}(2/5,-\tau '/5,-\tau /5)\qquad 
\pi ^{\perp }=\mathcal{A}(3/5,\tau '/5,\tau /5)
\end{equation} 
where  $\tau =(1+\sqrt{5})/2$, \ $\tau ' =(1-\sqrt{5})/2$ and
\begin{equation} 
\mathcal{A}(\alpha ,\beta ,\gamma )=\left( \begin{array}{rrrrr} 
\alpha &\beta &\gamma &\gamma &\beta  \\ 
\beta & \alpha &\beta &\gamma &\gamma \\  
\gamma &\beta & \alpha &\beta &\gamma \\ 
\gamma &\gamma &\beta &\alpha &\beta  \\ 
\beta &\gamma &\gamma &\beta &\alpha  
	     \end{array} \right) . 
\end{equation} 

Let $\kappa =1/\varrho =\sqrt{5/2}$, and let 
\begin{equation}
\mathbb{L}=\kappa \mathbb{Z}^5\qquad 
\mathbb{K}=v+\{ (x_1,x_2,x_3,x_4,x_5)\in \mathbb{R}^5 \ |\ 0\leq x_j\leq \kappa \} 
\end{equation}
where the translation vector $v\in E^\perp $ is chosen such that the boundary 
$\partial K$ of the set $K=\pi ^\perp (\mathbb{K})$ does not contain any element of 
$\pi ^\perp (\mathbb{L})$. This is possible since the set 
$\pi ^\perp (\mathbb{L})+\partial K$ has Lebesgue measure $0$.
The pattern defined in terms of the strip projection method
\begin{equation}
\mathcal{P}=\left\{ \left. \pi x\ \right| \ 
x\in \mathbb{L},\ \pi ^\perp x\in K\right\} 
\end{equation}
is the set of all the vertices of a {\em Penrose tiling} \cite{K}.

In order to study the self-similarities \cite{C1,Ma} of $\mathcal{P}$ it is convenient to
re-define it as a {\it multi-component model set} \cite{B}.
The subspace $E^\perp $ is the direct sum $E^\perp =E'\oplus E''$ of the 
$D_{10}$-invariant subspaces $E'$ and $E''$ corresponding to the orthogonal projectors 
\[ \pi '=\mathcal{A}( 2/5,-\tau /5,-\tau '/5)\qquad \pi ''=\mathcal{A}(1/5,1/5,1/5).\]
Let
\[ {\mathcal E}=E\oplus E'\qquad {\mathcal L}=(\pi +\pi ')(\mathbb{L})\qquad
L=\mathbb{L}\cap {\mathcal E}.\]
We have the relations
\begin{equation} 
E''=\{ (x_1,x_2,x_3,x_4,x_5)\in \mathbb{E}_5\ |\ x_1=x_2=x_3=x_4=x_5\}
\end{equation}
\begin{equation}  
{\mathcal E}=\{ (x_1,x_2,x_3,x_4,x_5)\in \mathbb{E}_5\ |\ x_1+x_2+x_3+x_4+x_5=0\}
\end{equation}
\begin{equation}  
{\mathcal L}=\mathbb{Z}w_1+\mathbb{Z}w_2+\mathbb{Z}w_3+\mathbb{Z}w_4
\end{equation}
where $w_j=(\pi +\pi ')(\kappa \varepsilon _j)$, that is, 
\begin{equation}\begin{array}{ll}
w_1=\frac{\kappa }{5} (4,-1,-1,-1,-1) \quad & 
w_2=\frac{\kappa }{5} (-1,4,-1,-1,-1) \\[2mm]
w_3=\frac{\kappa }{5} (-1,-1,4,-1,-1) \quad &
w_4=\frac{\kappa }{5} (-1,-1,-1,4,-1) .
\end{array}
\end{equation}

The matrices of the orthogonal projectors 
\begin{equation}
p:\mathcal{E}\longrightarrow \mathcal{E}:x\mapsto px=\pi x\qquad 
p':\mathcal{E}\longrightarrow \mathcal{E}:x\mapsto p'x=\pi 'x
\end{equation}
in the basis $\{ w_1,w_2,w_3,w_4\}$ are 
\begin{equation}\begin{array}{l}
p=\mathcal{B}\left( (5-\sqrt{5})/10,(5+\sqrt{5})/10,\sqrt{5}/5 \right)\\[2mm]
p'=\mathcal{B}\left( (5+\sqrt{5})/10,(5-\sqrt{5})/10,-\sqrt{5}/5 \right)
\end{array}
\end{equation}
where
\begin{equation} 
\mathcal{B}(\alpha ,\beta ,\gamma )=\left( \begin{array}{cccc} 
\alpha &\gamma & 0 & -\gamma  \\ 
0 &\beta & \gamma & -\gamma \\  
-\gamma &\gamma & \beta & 0 \\ 
-\gamma & 0 & \gamma &\alpha  
	     \end{array} \right) .
\end{equation} 
Since the restriction of $p$ to $\mathcal {L}$ is injective and
$\{ p'\,x \ |\ x\in \mathcal{L}\}=\pi '(\mathbb{L})$ is dense in $E'$ the 
collection of spaces and mappings
\begin{equation}\begin{array}{ccccccc}
p\, x \leftarrow x 
&:E& \stackrel{p}\longleftarrow & {\mathcal E}
& \stackrel{p'}\longrightarrow & E' :& x \rightarrow p'\, x\\
&&&\cup &&& \\
&&& {\mathcal L} &&&
\end{array}
\end{equation} 
is a cut and project scheme \cite{B,Ma,M}. We can assume that $v\in E'$, that is, $p'v=v$.

The lattice $\mathbb{L}$ is contained in the union 
$\bigcup_{n\in \mathbb{Z}}\mathcal{E}_n$ of the affine parallel subspaces
\begin{equation}
\fl {\mathcal E}_n=\{ (x_1,x_2,x_3,x_4,x_5)\in \mathbb{E}_5\ 
|\ x_1+x_2+x_3+x_4+x_5=n\kappa \}=h_n+\mathcal{E}
\end{equation}
where $h_n=(n\kappa ,0,0,0,0)\in \mathbb{L}$. 
Since $\mathbb{L}\cap \mathcal{E}_n=h_n+L$ the set 
\begin{equation}
{\mathcal L}_n=(\pi +\pi ')(\mathbb{L}\cap {\mathcal E}_n)=
(\pi +\pi ')h_n+L=nw_1+L
\end{equation}
is a coset of $L$ in ${\mathcal L}$ for any $n\in \mathbb{Z}$.
The set $\mathbb{K}\cap {\mathcal E}_n$ is non-empty only for $n\in \{ 0,1,2,3,4,5\}$,
but ${\mathcal K}_n=\pi '(\mathbb{K}\cap {\mathcal E}_n)$ has non-empty interior only for
$n\in \{ 1,2,3,4\}$. 
Let $\Omega \subset E'$ be the set of all the points lying inside or on the boundary of 
the regular pentagon with the vertices
\begin{equation}\begin{array}{l} 
\pi '(\kappa ,0,0,0,0)=\frac{\kappa }{5}\, (2,-\tau ,-\tau ',-\tau ', -\tau )\\[2mm]
\pi '(0,\kappa ,0,0,0)=\frac{\kappa }{5} (-\tau ,2,-\tau ,-\tau ',-\tau ')\\[2mm]
\pi '(0,0,\kappa ,0,0)=\frac{\kappa }{5} (-\tau ',-\tau ,2,-\tau ,-\tau ')\\[2mm]
\pi '(0,0,0,\kappa ,0)=\frac{\kappa }{5} (-\tau ',-\tau ',-\tau ,2,-\tau )\\[2mm]
\pi '(0,0,0,0,\kappa )=\frac{\kappa }{5} (-\tau ,-\tau ',-\tau ',-\tau ,2) .
\end{array}
\end{equation}
One can remark that $\mathcal{K}_1=v+\Omega $, 
${\mathcal K}_2=v-\tau \Omega $, 
${\mathcal K}_3=v+\tau \Omega $,
${\mathcal K}_4=v-\Omega$,  
and one can re-define the pattern $\mathcal{P}$ 
as a multi-component model set
\begin{equation}
\mathcal{P}=\bigcup_{n=1}^4
\left\{ p\, x\ \left| \ x\in {\mathcal L}_n , \ p'\, x\in {\mathcal K}_n
\right.\right\} .
\end{equation}
This definition is directly related to de Bruijn's definition \cite{M}.

The matrix of the linear transformation $S=\lambda \, p +\lambda '\, p'$ 
is $\mathcal{B}(k,l,m)$, where 
\begin{equation} \fl
k=\lambda \frac{5-\sqrt{5}}{10}+\lambda '\frac{5+\sqrt{5}}{10}\qquad
l=\lambda \frac{5+\sqrt{5}}{10}+\lambda '\frac{5-\sqrt{5}}{10}\qquad
m=\lambda \frac{\sqrt{5}}{5}-\lambda '\frac{\sqrt{5}}{5}
\end{equation}
and it has integer entries if and only if 
$\lambda =k+m\tau $ and $\lambda '=k+m\tau '$ with $k,m\in \mathbb{Z}$.

Let $\lambda =k+m\tau $ be a fixed element belonging to the infinite set
\begin{equation} 
\Lambda =\{ k+m\tau \ |\ k,m\in \mathbb{Z},\ 2m-k+1\in 5\mathbb{Z},
\ |k+m\tau '|<1/2\}
\end{equation}
and let $\lambda '=k+m\tau '$. The matrix of transformation
$\tilde{S}_\lambda : \mathbb{E}_5\longrightarrow \mathbb{E}_5$, \ 
$\tilde{S}_\lambda =\lambda \pi +\lambda '\pi '+\pi ''$
in basis $\{ \varepsilon _1,\varepsilon _2,\varepsilon _3,
\varepsilon _4,\varepsilon _5\}$ is
\[ \mathcal{A}\left( k+\frac{2m-k+1}{5},\, \frac{2m-k+1}{5},\,
\frac{2m-k+1}{5}-m\right) .\]
Since $\tilde{S}_\lambda (\mathcal{E}_n)\subset \mathcal{E}_n$ and the
matrix of $\tilde{S}_\lambda $ has integer entries we get 
$\tilde{S}_\lambda (\mathbb{L}\cap \mathcal{E}_n)\subset \mathbb{L}\cap \mathcal{E}_n$,
whence $S_\lambda (\mathcal{L}_n)\subset \mathcal{L}_n$, for any $n\in \mathbb{Z}.$ 

From $|k+m\tau '|<1/2$ it follows
$\lambda '(\mathcal{K}_n-v)+v\subset \mathcal{K}_n$ \
for any $n\in \{ 1,2,3,4\}.$ More than that, there exists $\delta >0$ such that
\begin{equation}
\left. \begin{array}{r}
v'\in E'\\
||v'-v||<\delta 
\end{array}\right\}\quad \Longrightarrow \quad 
\lambda '(\mathcal{K}_n-v')+v'\subset \mathcal{K}_n
\end{equation}
for any $n\in \{ 1,2,3,4\}.$ Since $p'(L)$ is dense in $E'$ the set
\begin{equation}
 \mathcal{Q}_\lambda =\{ y\in L\ |\ ||p'y-v||<\delta \} 
\end{equation}
is an infinite set. For each $\lambda \in \Lambda $ and for each 
$y\in \mathcal{Q}_\lambda $ we have
\begin{equation} \fl
\left. \begin{array}{r}
x\in \mathcal{L}_n\\
p'x\in \mathcal{K}_n
\end{array} \right\}\quad \Longrightarrow \quad 
\left\{ \begin{array}{l}
S_\lambda (x-y)+y\in \mathcal{L}_n\\
p'[S_\lambda (x-y)+y]=\lambda '(p'x-p'y)+p'y\in \mathcal{K}_n
\end{array} \right.
\end{equation}
whence
\begin{equation}
  px\in \mathcal{P}\quad \Longrightarrow \quad 
p[S_\lambda (x-y)+y]=\lambda (px-py)+py\in \mathcal{P} .
\end{equation}
This means that $\mathcal{P}$ is invariant under the self-similarity
\begin{equation}
 E\longrightarrow E: z\mapsto \lambda (z-py)+py =\lambda z+(1-\lambda )py
\end{equation}
of center $py$ and scaling factor $\lambda $. The Penrose tiling 
is transformed into a similar tiling inflated by $\lambda $ with vertices
belonging to $\mathcal{P}$.


\section*{References}

\begin{thebibliography}{10}
\bibitem{B} 
Baake M and Moody R V 1999 Multi-component model sets and invariant
          densities {\it Proc. Int. Conf. Aperiodic' 97 
          (Alpe d'Huez, 27-31 August, 1997)}
          ed M de Boissieu {\it et al.} (Singapore: World Scientific)
          pp 9-20
\bibitem{C1}
Cotfas N  1998 On the self-similarities of the three-dimensional Penrose pattern
          {\it J. Phys. A: Math. Gen.} {\bf 31} 7273-7277
\bibitem{C2}
Cotfas N 1999 Permutation representations defined by $G$-clusters with
         application to quasicrystals {\it Lett. Math. Phys.} {\bf 47} 111-23
\bibitem{K}
Katz A and Duneau M 1986 Quasiperiodic patterns and icosahedral symmetry 
          {\it J. Phys. (France)} {\bf 47} 181-196
\bibitem{Ma}
Mas\' akov\' a Z, Patera J and Pelantov\' a 1998 Inflation centers of the cut
          and project quasicrystals {\it J. Phys. A: Math. Gen.} {\bf 31} 1443-53
\bibitem{M} 
Moody R V Meyer sets and their duals {\it The Mathematics of Long-Range
        Aperiodic Order} ed R V Moody (Dordrecht: Kluwer) pp 411-12
\end{thebibliography}
\end{document}